\documentclass{PoS}

\title{Pre-maximum and maximum of Novae: The spectroscopic observations of Nova ASASSN-17hx}

\ShortTitle{Pre-maximum and maximum of Novae: Nova ASASSN-17hx}

\author{\speaker{Rosa Poggiani}\\
        Universit\`a di Pisa and Istituto Nazionale di Fisica Nucleare, Sezione di Pisa\\
        E-mail: \email{rosa.poggiani@df.unipi.it}}

\abstract{The coverage of the  pre-maximum stage of novae is sparse, with the exception of a few slow novae where the pre-maximum duration is of the order of some weeks. This paper discusses the main features of novae during the pre-maximum and the maximum stages and reports the preliminary results of an investigation of ASASSN-17hx, a peculiar nova that showed a long pre-maximum and secondary brightenings.}

\FullConference{The Golden Age of Cataclysmic Variables and Related Objects IV\\
		11-16 September, 2017\\
		Palermo, Italy}

\begin{document}

\section{Introduction}

Classical novae \cite{PayneGaposchkin1957}, \cite{BodeEvans2008} are cataclysmic variables where a white dwarf accretes material from a normal star. The accreted material accumulates on the surface of the white dwarf, undergoing compression, until the increase in temperature and density produces a thermonuclear runaway \cite{BodeEvans2008}. The rise to the maximum brightness can be as short as some hours, making the observations of novae during the pre-maximum  stage difficult. There are a few notable exceptions, the slow novae HR Del, V1548 Aql, V723 Cas, V5558 Sgr, that showed pre-maxima stages lasting for some months. A recent peculiar  nova, ASASSN-17hx, showed a long pre-maximum and secondary maxima after the main outburst. The present paper discusses the pre-maximum and the maximum stages of novae and presents the preliminary  results  of the spectroscopic monitoring of ASASSN-17hx performed by the author.

\section{Novae: the pre-maximum and maximum stages}

The main features of novae during the pre-maximum and maximum stages have been summarized by \cite{BodeEvans2008}. The evolution of a nova towards the maximum brightness includes an initial brightening stopping at about two magnitudes from the maximum, followed by a pre-maximum plateau and by the final brightening to the maximum. The time scales of the stages are different: the initial brightening occurs within one day, while the pre-maximum stage can last from hours to months and the maximum stage duration ranges from hours to a few days. The duration of the pre-maximum stage is related to the speed class of nova, being of the order of hours in fast novae and from days to months in slow novae. The spectra during the pre-maximum stage show broad and blue shifted absorptions and P Cyg profiles. The spectral type at maximum ranges from B to F, where later types are related to the slower novae.  The pre-maximum spectra are explained by a uniformly expanding and optically thick envelope that undergoes cooling during the expansion.

\begin{figure}
\centering
\includegraphics[width=.35\textwidth]{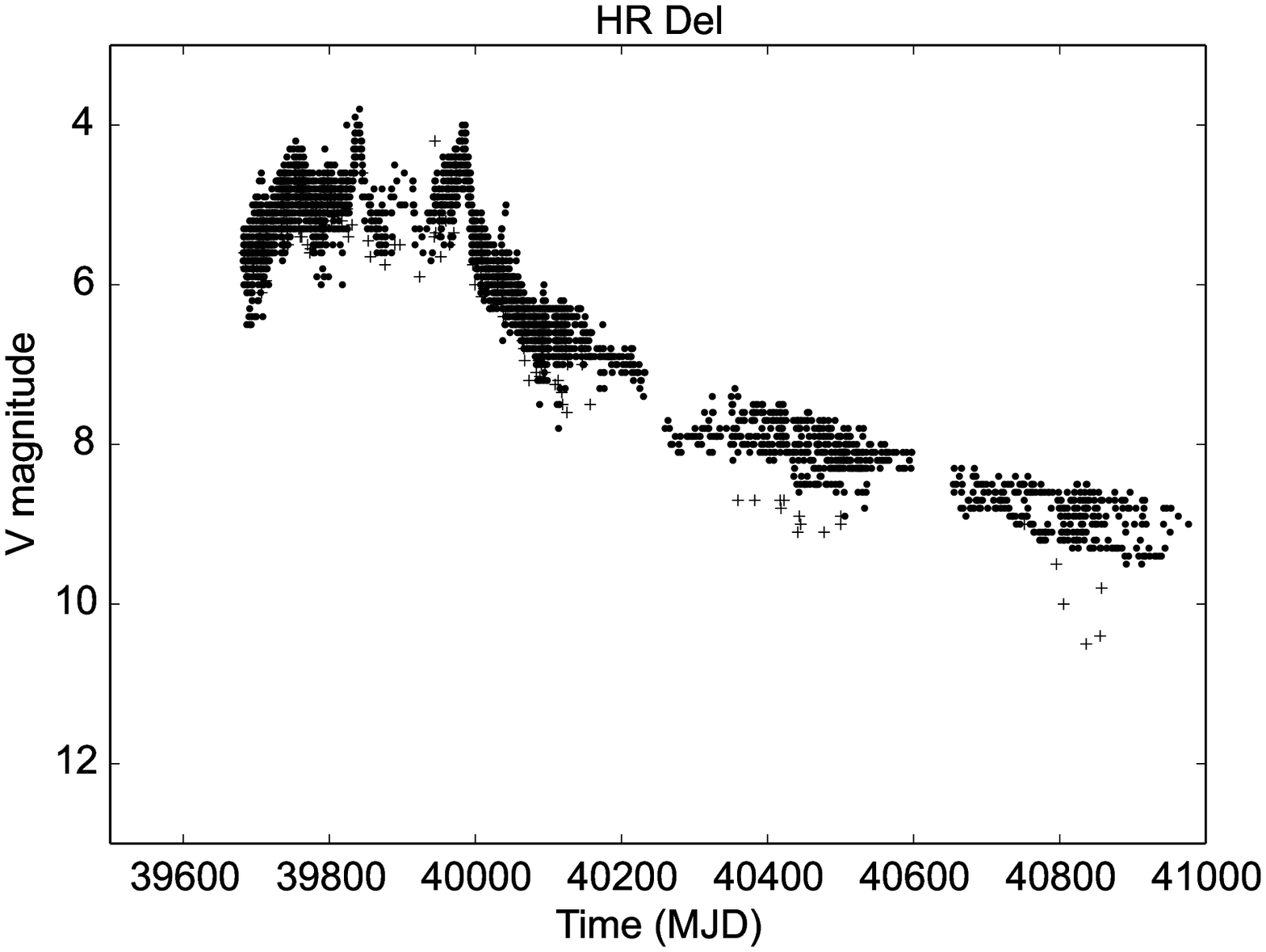}
\includegraphics[width=.35\textwidth]{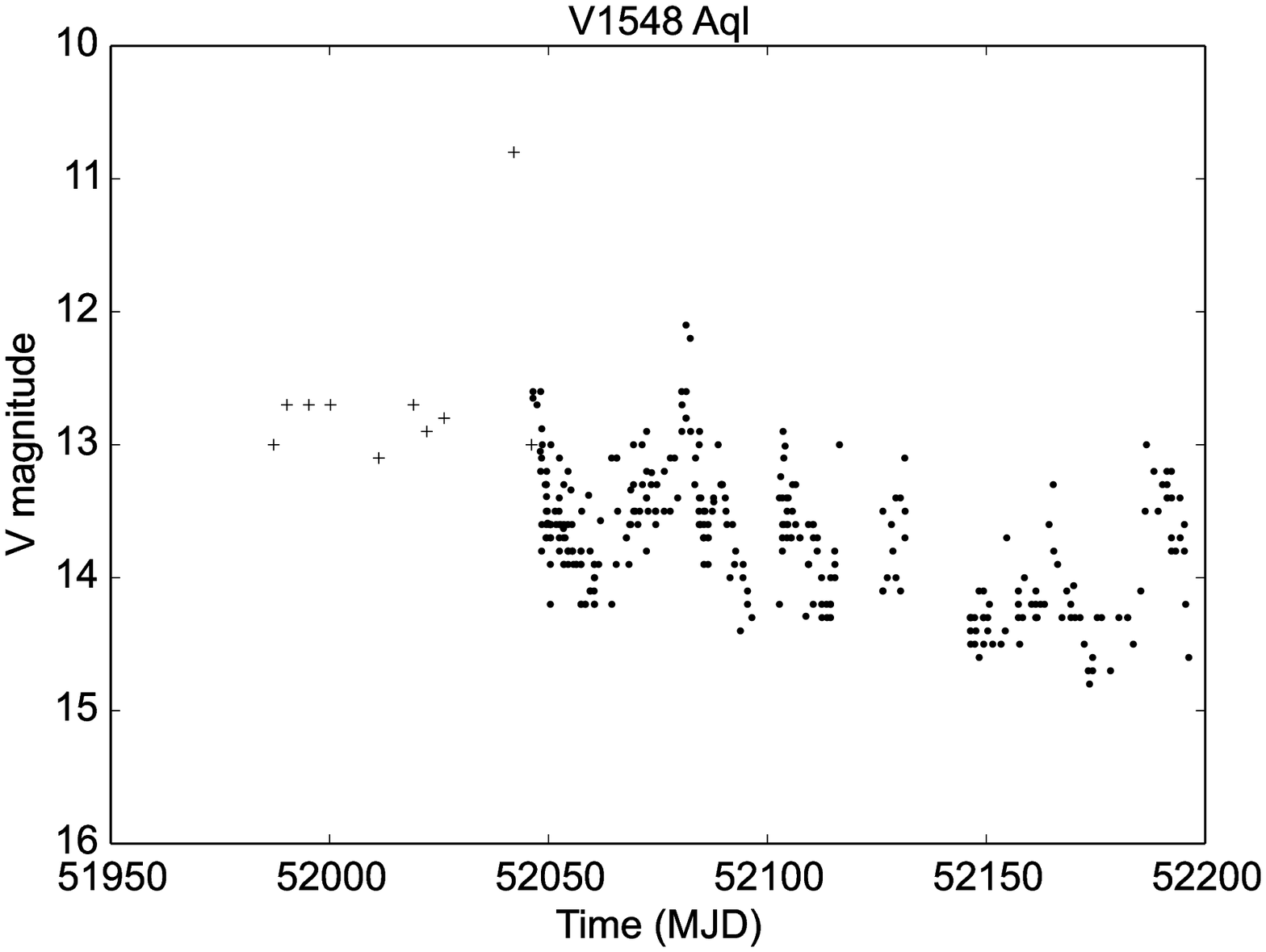}
\includegraphics[width=.35\textwidth]{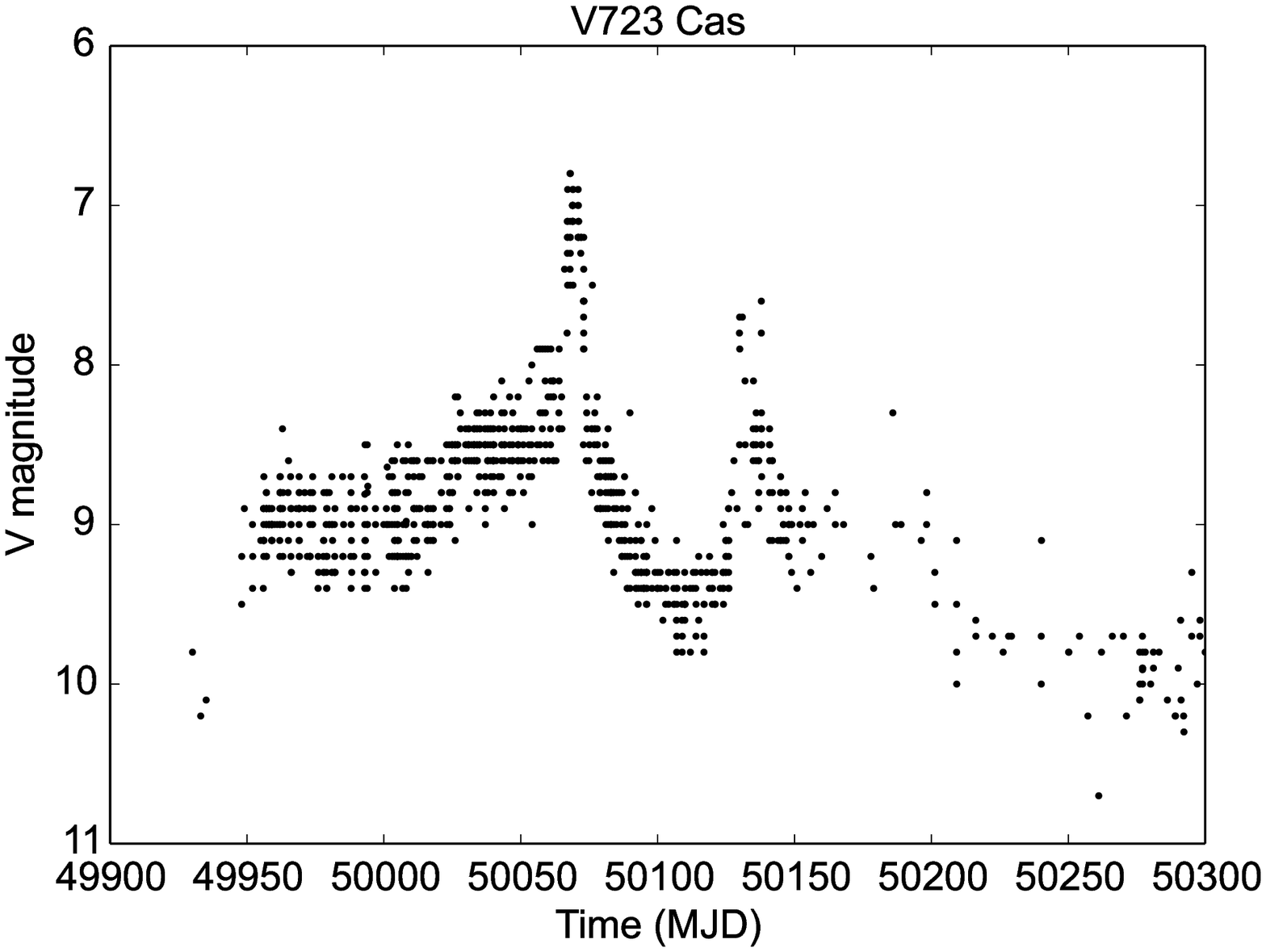}
\includegraphics[width=.35\textwidth]{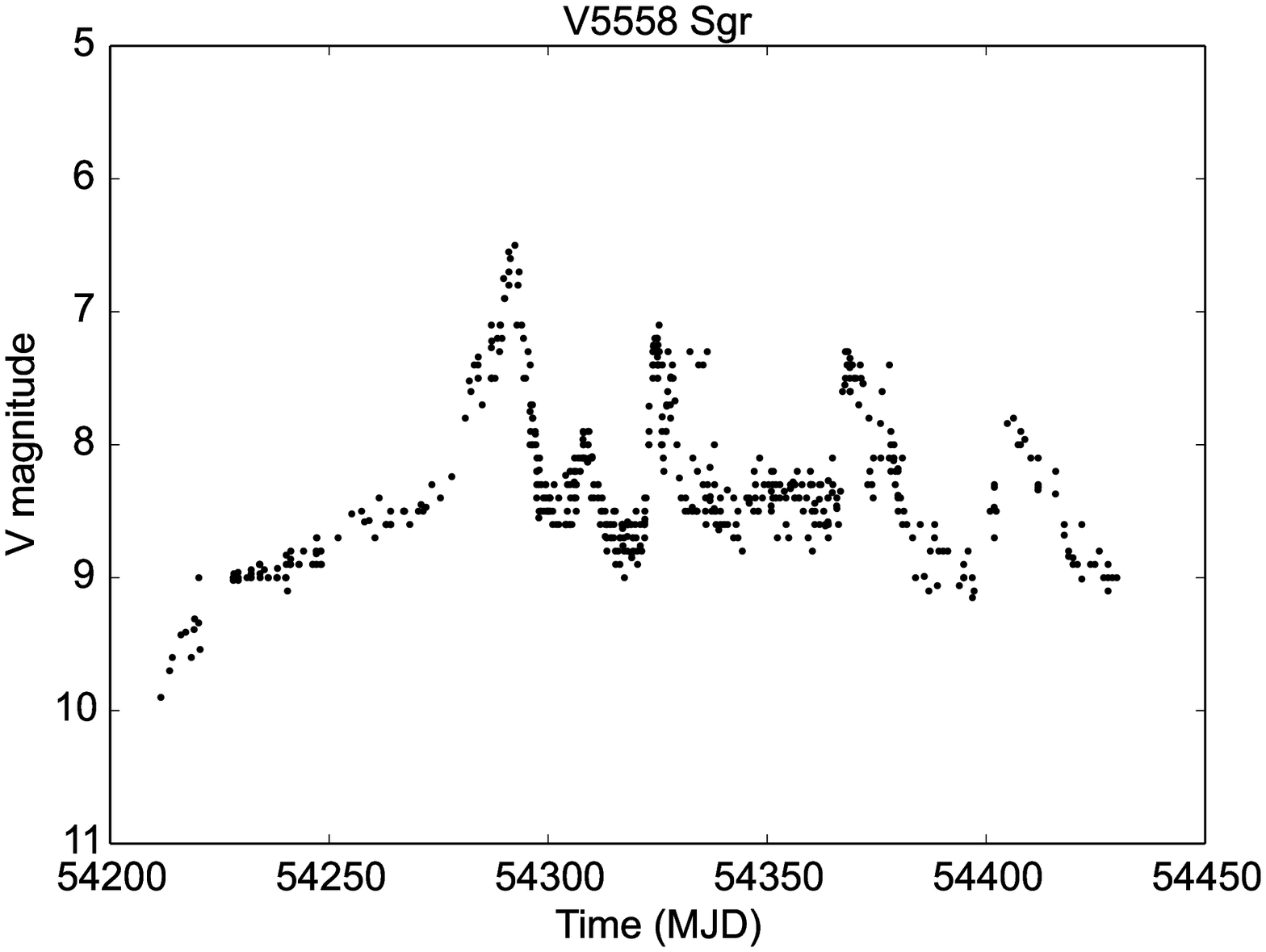}
\caption{Light curves of the slow novae HR Del, V1548 Aql, V723 Cas, V5558 Sgr; data of VSNET and AFOEV archives}
\label{fig:preslow}
\end{figure}

A few novae have been observed during the pre-maximum stage: the slow novae V1548 Aql \cite{Kato2001}, HR Del \cite{Friedjung1992}, V723 Cas \cite{Iijima1998}, V5558 Sgr \cite{Poggiani2008}, \cite{Tanaka2011b} and the fast nova V463 Sct \cite{Kato2002}. The light curves of V1548 Aql, HR Del, V723 Cas, V5558 Sgr, built using the VSNET \footnote{http://ooruri.kusastro.kyoto-u.ac.jp/mailman/listinfo/} and AFOEV \footnote{http://cdsarc.u-strasbg.fr/afoev/} data, are reported in Fig. \ref{fig:preslow}. All slow novae show multiple rebrightnenings during the decline after the maximum.
The white dwarfs of slow novae have masses close to 0.6 M$_{\odot}$, just at the limit of thermonuclear runaway \cite{Friedjung1992}.

\begin{figure}
\centering
\includegraphics[width=.35\textwidth]{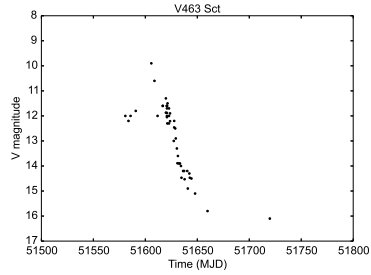}
\caption{Light curve of nova V463 Sct; data of VSNET and AFOEV archives}
\label{fig:prefast}
\end{figure}

There is a single exception of fast nova with a long pre-maximum stage, V463 Sct (Fig. \ref{fig:prefast}). The object shows that long pre-maximum stages are not necessarily associated to slow novae.

The pre-maximum stage of novae belonging to both speed classes has been discussed by \cite{Hachisu2004}. A model with a steady state optically thick wind reproduced the light curves of the slow nova V723 Cas and of the fast nova V463 Sct. When the hydrogen rich envelope is massive enough, expansion occurs with a temperature drop, mimicking the behavior of a supergiant. The visual magnitude is close to bolometric magnitude for months, producing the pre-maximum halt, whose duration depends on the envelope mass \cite{Hachisu2004}. The absolute magnitude of the pre-maximum acts as a standard candle of the Eddington luminosity. The authors estimated a mass of 0.59 M$_{\odot}$ for V723 Cas and of 1.1 M$_{\odot}$ for V463 Sct, in agreement with their speed class. 

The evolution of novae with low mass dwarfs (0.65 M$_{\odot}$) before the eruption and during the decline has been discussed by \cite{Hillman2014} in the context of visual, near UV and X-ray emission. The model light curves show a fast rise and a smoother decline of the near UV luminosity, before achieving the maximum optical luminosity. The precursor UV flashes, with a luminosity increase up to a few magnitudes, last from hours to a few days and are potentially observable by high cadence monitoring. The UV flashes have been observed in M31 2009-10b \cite{Cao2012}, M31 2010-11a \cite{Cao2012}, M31 2007-07c \cite{Pietsch2007a}, M31 2007-07d \cite{Pietsch2007b}. The model discusses the dips sometimes observed during the pre-maximum stage. Since the luminosity follows the effective temperature, the dips correspond to temperature drops before the onset of mass loss. The energy flux drops when convection in the expanding envelope  becomes inefficient near the envelope surface and opacity decreases. Dips have been observed in RS Oph and KT Eri \cite{Hounsell2010} and in V5589 Sgr (see below) \cite{Eyres2017}.

The present all sky surveys are providing high cadence observations of novae during the pre-maximum stage, offering the opportunity to test models.
The space based Solar Mass Ejection Imager (SMEI) instrument performs photometric observations with a cadence of about 100 minutes. The instrument has monitored 13 galactic novae \cite{Hounsell2010}, \cite{Hounsell2016}, some of them during the pre-maximum. The photometric evolution of the pre-maximum of V5589 Sgr has been investigated by \cite{Eyres2017} with a cadence of 80 min using STEREO  HI-1B. The authors reported evidence of pre-maximum halts, detecting two short duration dips during the rise to maximum, as predicted by \cite{Hillman2014}. The halts are precursors of the onset of mass loss \cite{Eyres2017}. The data of V5589 Sgr have been re-analyzed by \cite{Thompson2017}, who performed a different background subtraction. The authors derived a smoother light curve, with no evidence of dips. 

Novae can be classified into two spectral classes, the Fe II and He/N class, according to the strongest lines in the post-outburst spectra, besides the Balmer ones \cite{Williams1991}, \cite{Williams1992}, \cite{Williams1994}. He/N novae exhibit high ejection velocities and fast declines, while Fe II novae show low to high ejection velocities and either fast or slow declines.
The classification by \cite{Williams1991}, \cite{Williams1992}, \cite{Williams1994} is based on post-outburst spectra and not on pre-maximum spectra. The switching between different classes during the pre-maximum has been observed in a few novae. During the early pre-maximum of V5558 Sgr \cite{Tanaka2011b} and T Pyx \cite{Imamura2012}, \cite{Ederoclite2014} He  I  lines were observed before Fe II appearance. He/N and Fe II novae spectra are explained by the dwarf ejecta and by a circumbinary gas envelope, respectively \cite{Williams2012}. The relative contribution of the two mechanisms varies during the evolution \cite{Williams2012}.

\section{Nova ASASSN-17hx}

Nova ASASSN-17hx was discovered on 2017 June 23 by ASAS-SN, the All-Sky Automated Survey for Supernovae, during the  rise to maximum \cite{Stanek2017}. ASAS-SN is a system of telescopes observing 20000 square degrees per night \cite{asas}. ASASSN-1 (Brutus) at Haleakala and ASASSN-2 (Cassius) at Cerro Tololo include four robotic 14 cm telescopes each.
The progenitor of ASASSN-17hx was identified by \cite{Saito2017}. Spectroscopic investigations were performed by different authors. Balmer and He I lines were observed by \cite{Kurtenkov2017} on June 24; the lines showed P Cyg profiles on June 29 \cite{Williams2017}, suggesting that ASASSN-17hx could be classified as a He/N nova. After June 29 the He I lines weakened and Fe II lines appeared \cite{Berardi2017}. An apparent maximum was observed by \cite{Munari2017a} on July 10, when the nova showed Balmer and Fe II lines with P Cyg profiles, suggesting that it could be a Fe II nova. The Balmer and Fe II lines with P Cyg profiles were still present on July 3 \cite{Pavana2017}. ASASSN-17hx was detected by the UVOT telescope of Swift, but not by the XRT instrument \cite{Kuin2017}.
During the decline after the main maximum, the spectra of ASASSN-17hx showed Fe II and He I lines and Balmer lines with P Cyg profiles \cite{Munari2017b}. ASASSN-17hx has shown a first  rebrightening \cite{Munari2017c}, \cite{Kurtenkov2017b}, with spectra showing Balmer and Fe II lines with P Cyg absorptions, as in Fe II novae, before the reappearance of He I \cite{Guarro2017}.

\begin{figure}
\centering
\includegraphics[width=.6\textwidth]{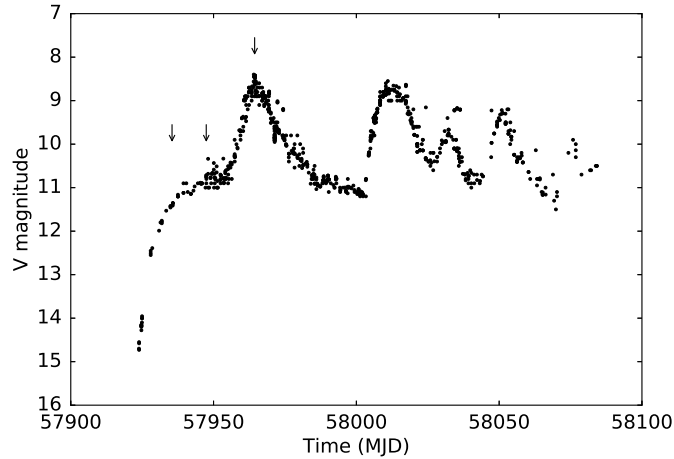}
\caption{Light curve of nova ASASSN-17hx; data of VSNET, AAVSO and ASAS archives}
\label{fig:asas}
\end{figure}

The light curve of ASASSN-17hx is reported in Fig. \ref{fig:asas}; the curve has been built using the data of VSNET, AAVSO \footnote{https://www.aavso.org/}, ASAS \footnote{http://www.astronomy.ohio-state.edu/asassn/index.shtml}. The arrows mark the epochs of the spectroscopic observations discussed in the present paper, that sample the rise to the plateu before maximum, the plateau stage and the maximum. The full results of the monitoring of ASASSN-17hx will be the subject of a forthcoming paper. 

ASASSN-17hx shows multiple rebrightenings after the main maximum; at the epoch of the conference, the luminosity of ASASSN-17hx was rising towards the first rebrightening, whose amplitude approaches the amplitude of the main maximum. The estimated epoch of maximum is MJD=57964.5, when ASASSN-17hx achieved a magnitude of 8.4. The apparent decline time by two magnitudes is 17$\pm$1 day. The light curve of ASASSN-17hx is similar to the curves of HR Del, V723 Cas, V5558 Sgr of Fig. \ref{fig:preslow}, thus it is a slow nova, from the point of view of the photometric evolution. ASASSN-17hx probably harbors a 0.6 M$_{\odot}$ white dwarf, just at the limit of thermonuclear runaway \cite{Friedjung1992}, \cite{Prialnik1995}, as other slow novae. Another possibility is the local thermonuclear runaway discussed by \cite{Orio1993}. An independent estimation of the white dwarf mass is necessary to discriminate between the models. 

I have monitored ASASSN-17hx during the pre-maximum and maximum stages at the Cassini 1.5m telescope, Loiano Observatory, equipped with the BFOSC Imager/Spectrograph.
Most spectra have been secured with grisms \#4 (range 3800-8700~\AA, resolution 3.97~\AA/pixel), \#8 (range 6100-8180~\AA, resolution 1.60~\AA/pixel). The observations are part of a program of monitoring of Northern novae \cite{Poggiani2012}.

\begin{figure}
\centering
\includegraphics[width=.45\textwidth]{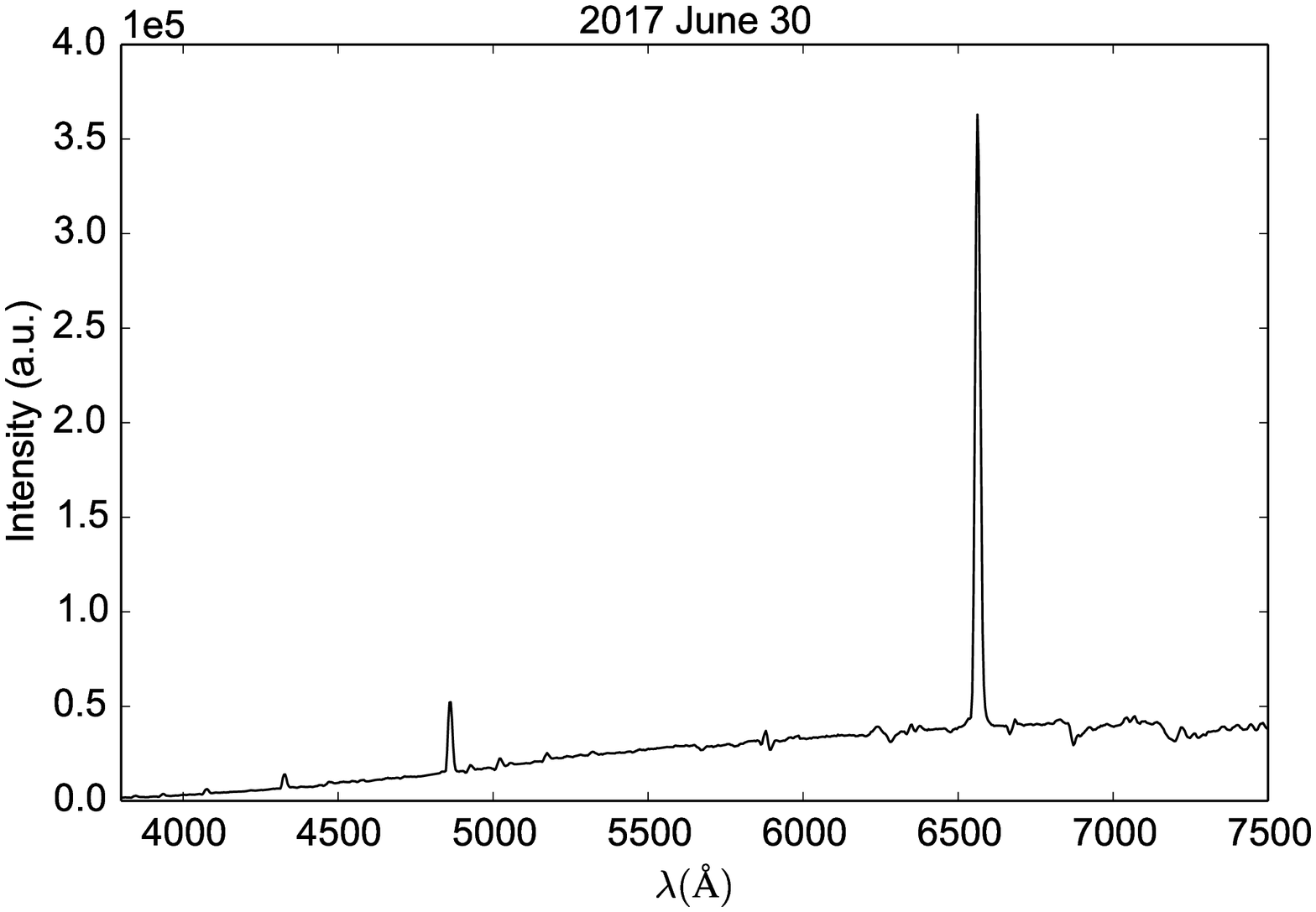}
\includegraphics[width=.45\textwidth]{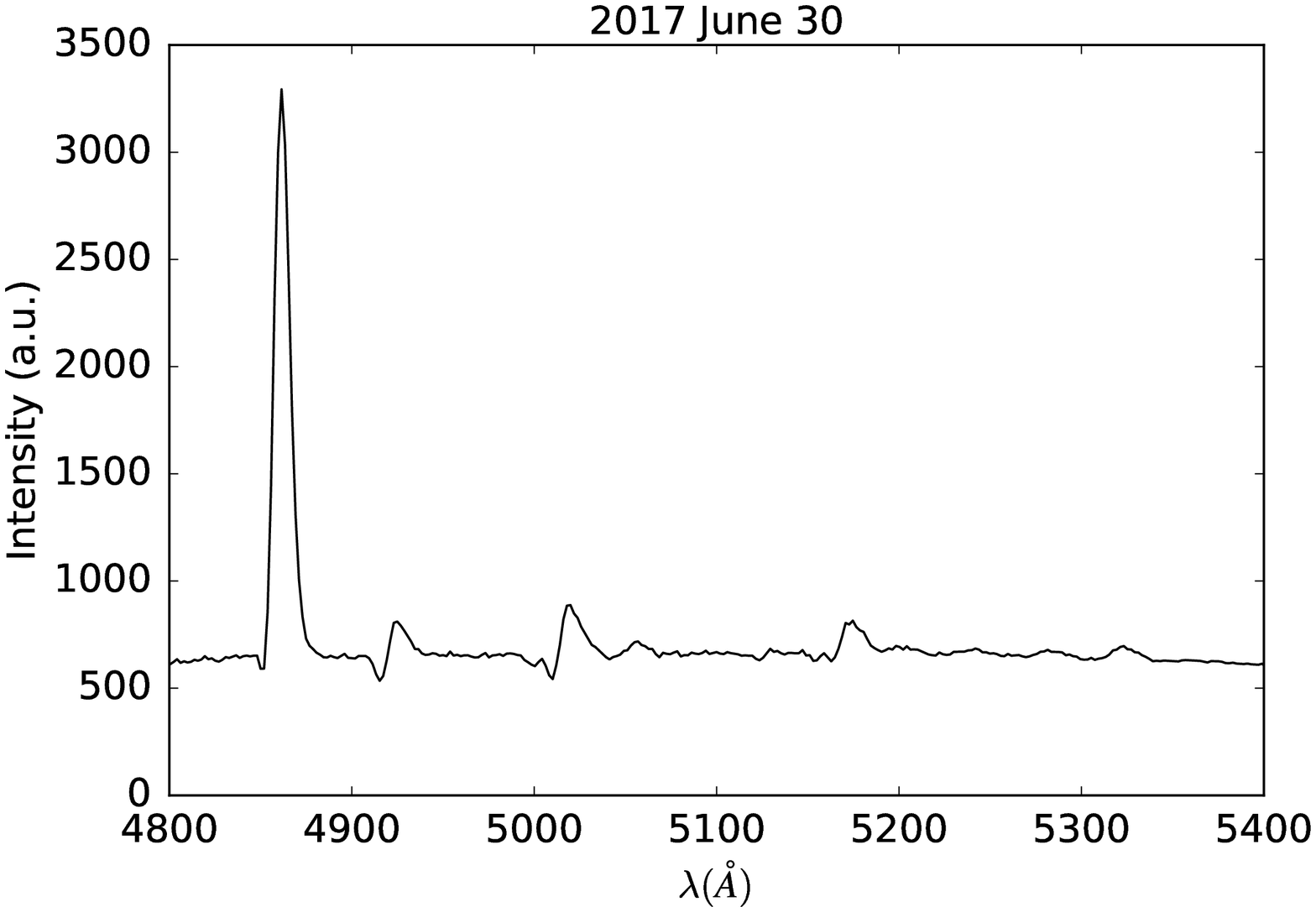}
\caption{Left: spectrum of ASASSN-17hx secured during the rise to the plateau; right: the H$\beta$, Fe II (42) region}
\label{fig:rise}
\end{figure}

A spectrum secured on 2017 June 30, during the rise to  the  plateau  before  maximum, showed lines of different transitions: Balmer (H$\alpha$, H$\beta$, H$\gamma$), Fe II (42) (4924, 5018, 5169~\AA), Fe II (37) and Fe II (38) (around 4500~\AA), He I 5876, Si II 6347, 6371 (Fig. \ref{fig:rise}, left). The lines show P Cyg profiles (Fig. \ref{fig:rise}, right).The spectrum is similar to the spectra of V723 Cas \cite{Iijima1998} and V5558 Sgr \cite{Tanaka2011b} at the same stage.

\begin{figure}
\centering
\includegraphics[width=.45\textwidth]{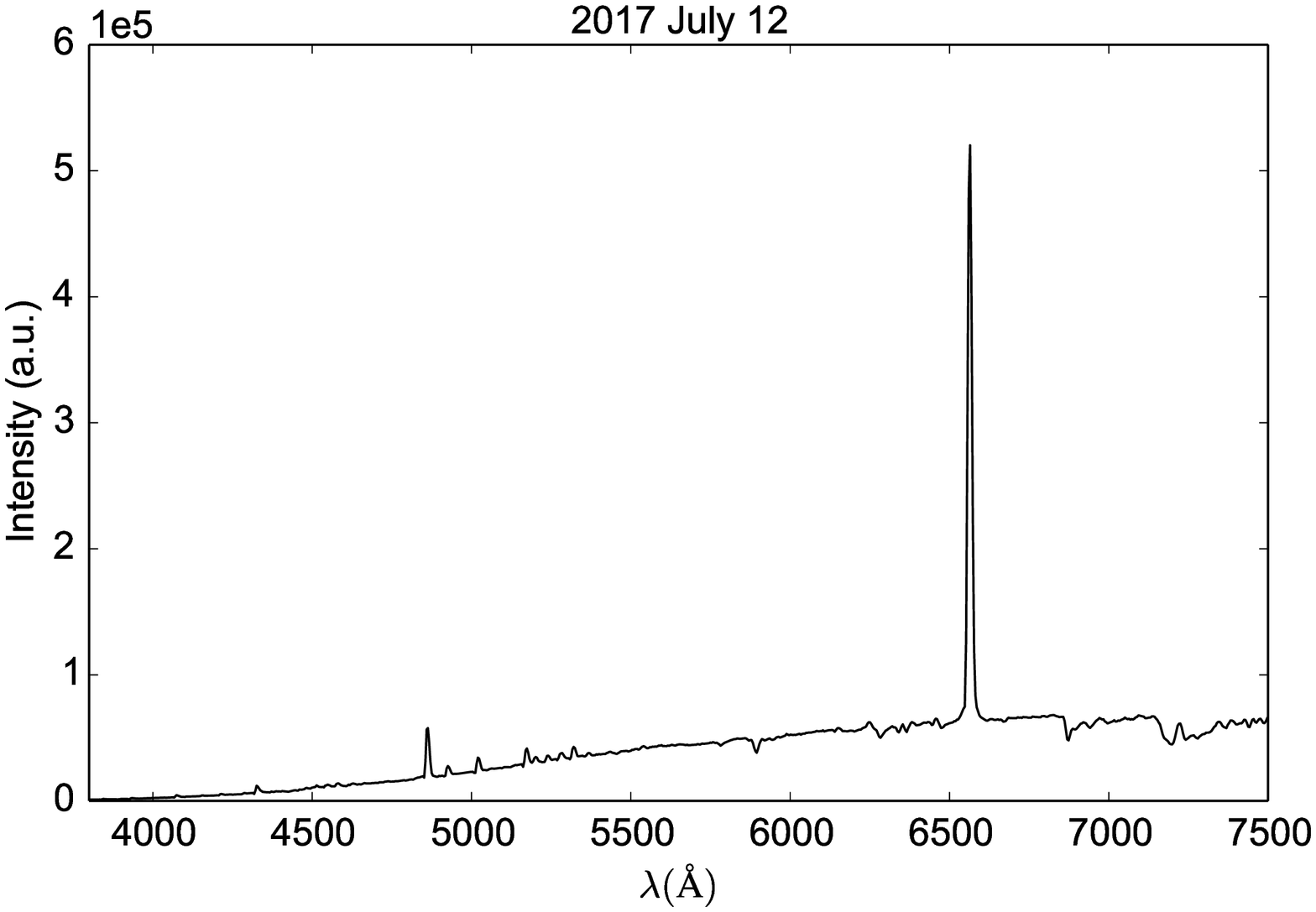}
\includegraphics[width=.45\textwidth]{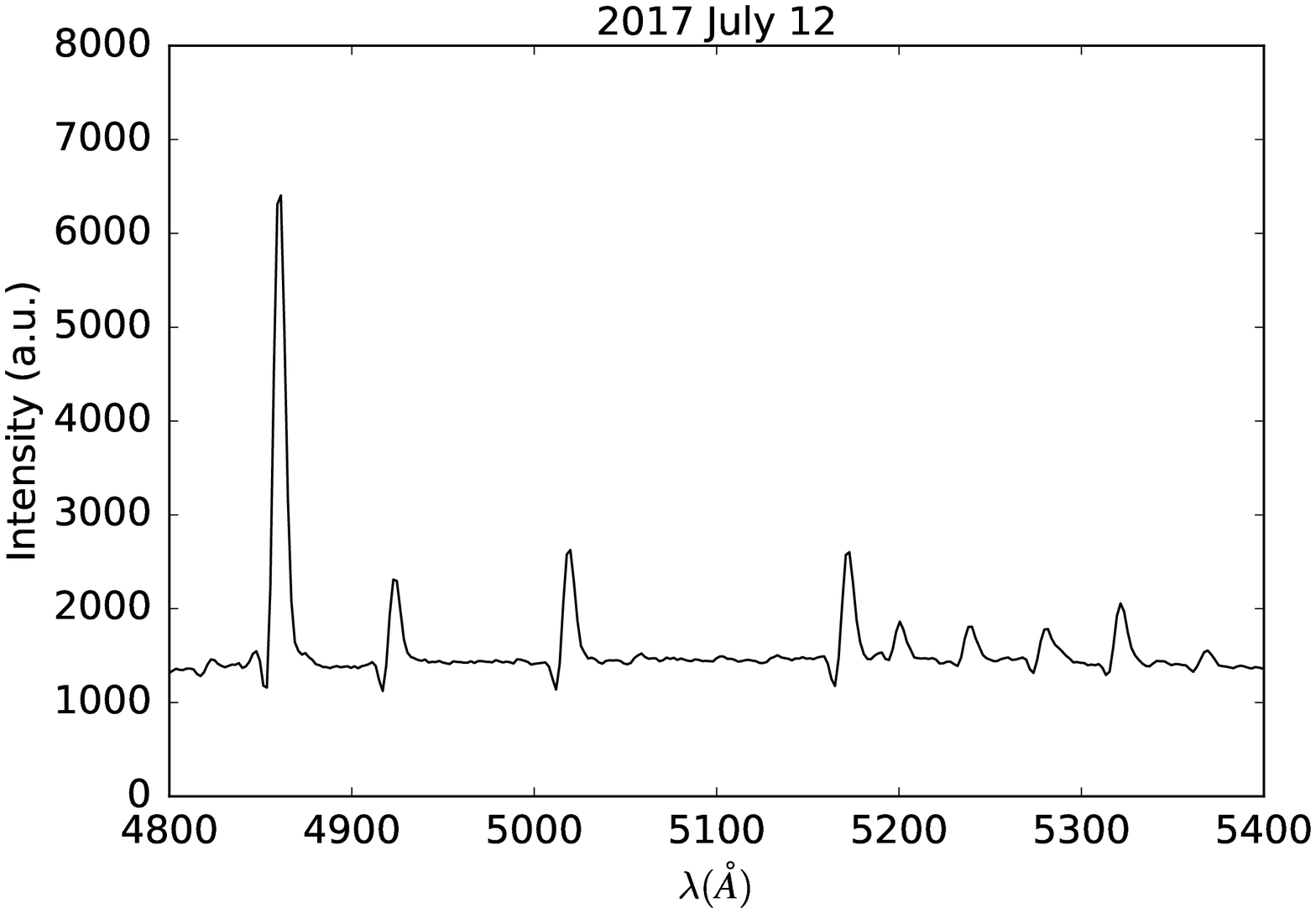}
\caption{Left: spectrum of ASASSN-17hx secured during the pre-maximum plateau; right: the H$\beta$, Fe II (42) region}
\label{fig:plateau}
\end{figure}

A spectrum secured on July 12 during the plateau stage showed an increasing intensity of Fe II (42) lines and the emerging of Fe II (48), Fe II (49) lines around 5300~\AA (Fig. \ref{fig:plateau}, left). The absorption component of P Cyg profiles of the lines has increased (Fig. \ref{fig:plateau}, right). The spectrum is similar to the spectra of V723 Cas \cite{Iijima1998} and V5558 Sgr \cite{Poggiani2008}, \cite{Tanaka2011b} during the plateau stage.

\begin{figure}
\centering
\includegraphics[width=.45\textwidth]{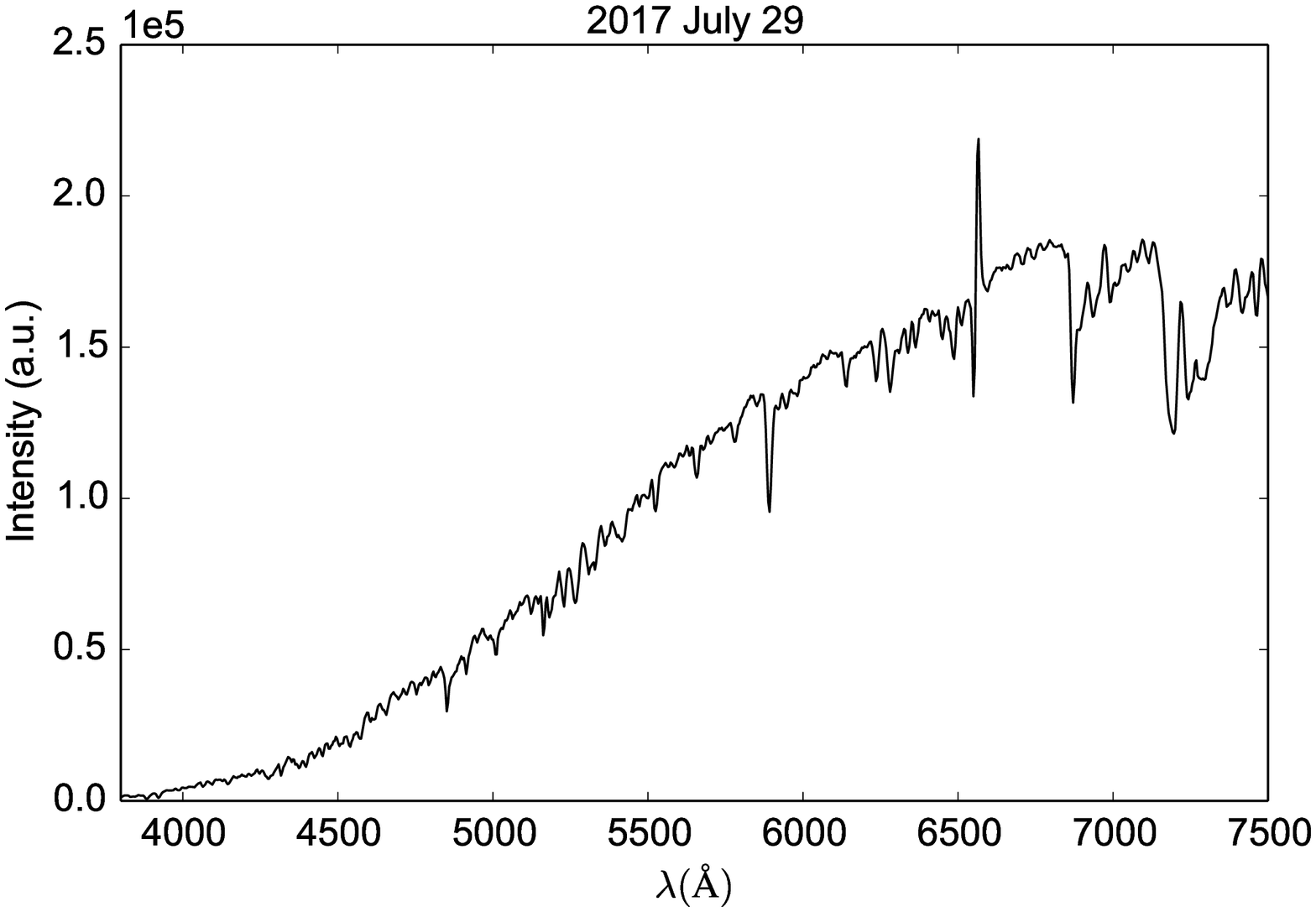}
\includegraphics[width=.45\textwidth]{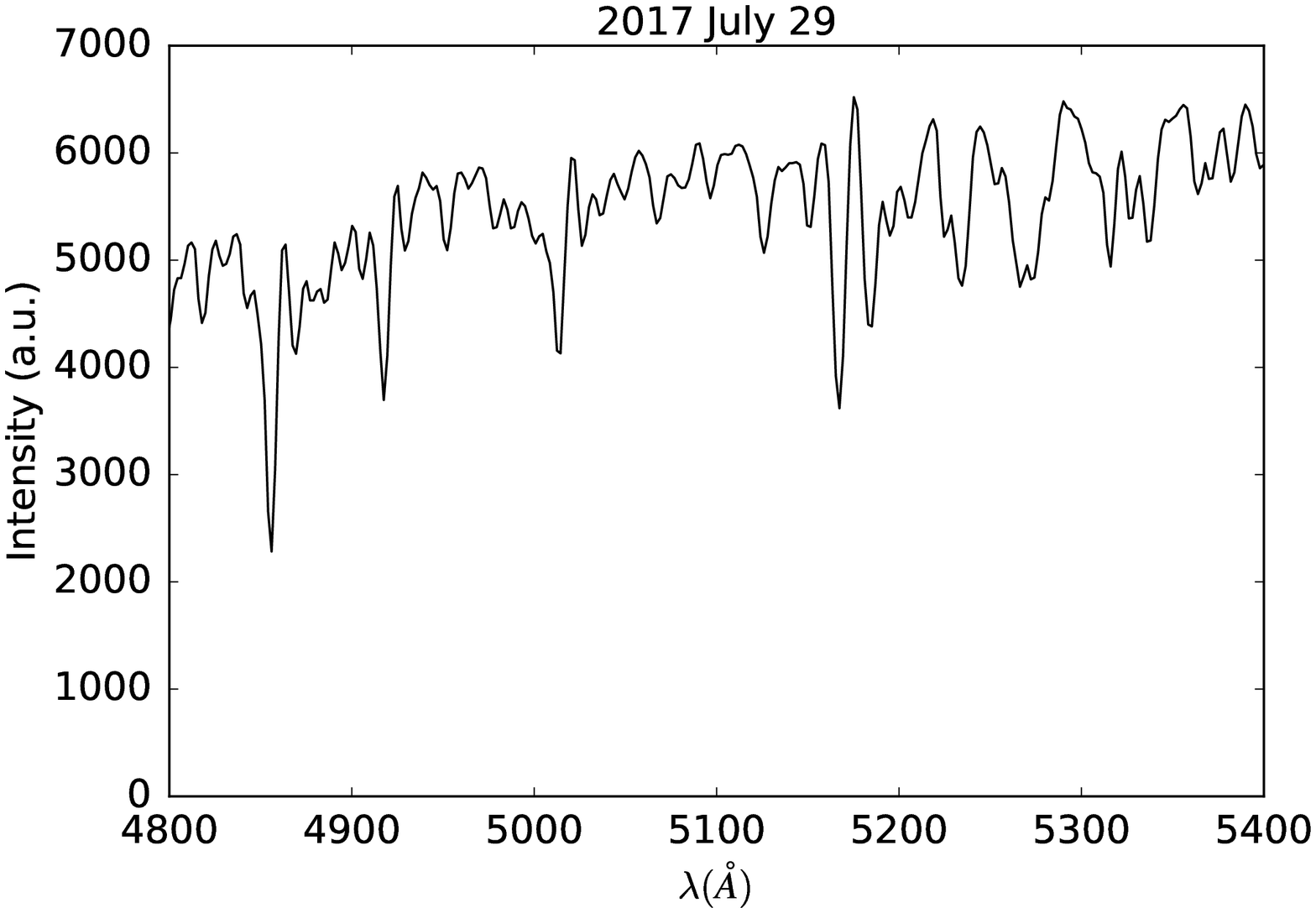}
\caption{Left: spectrum of ASASSN-17hx secured around the maximum; right: the H$\beta$, Fe II (42) region}
\label{fig:maximum}
\end{figure}

The spectrum of ASASSN-17hx secured around the maximum, on July 29 shows the H$\alpha$ transition with a strong emission component (Fig. \ref{fig:maximum}, left). On the other hand, H$\beta$ and Fe II lines show prominent absorptions (Fig. \ref{fig:maximum}, right). The spectrum is similar to the spectrum of V723 Cas \cite{Iijima1998} and V5558 Sgr at the same stage \cite{Tanaka2011b}, confirming that novae at maximum show a supergiant like spectrum with H$\alpha$ in emission.

\begin{figure}
\centering
\includegraphics[width=.45\textwidth]{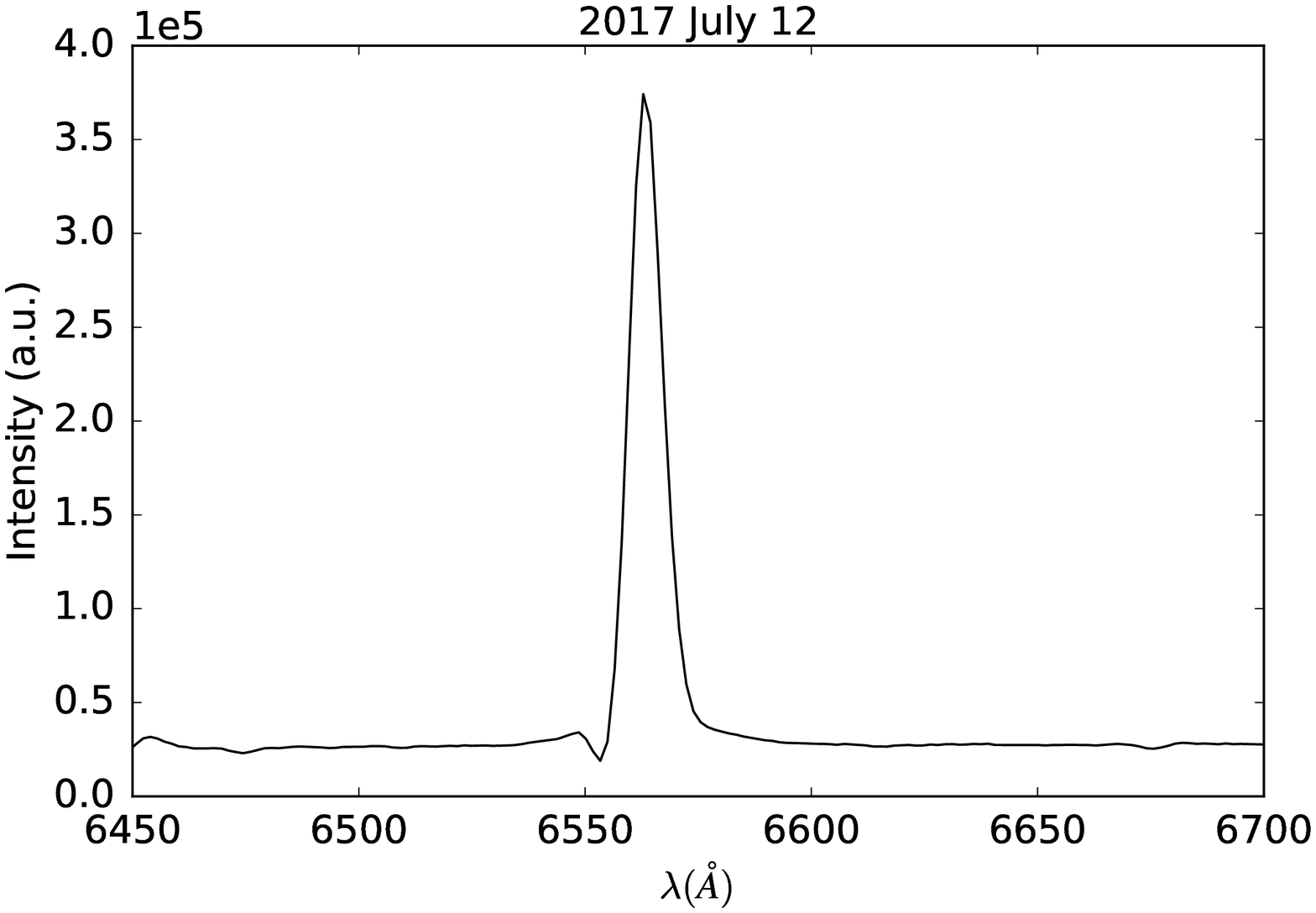}
\includegraphics[width=.45\textwidth]{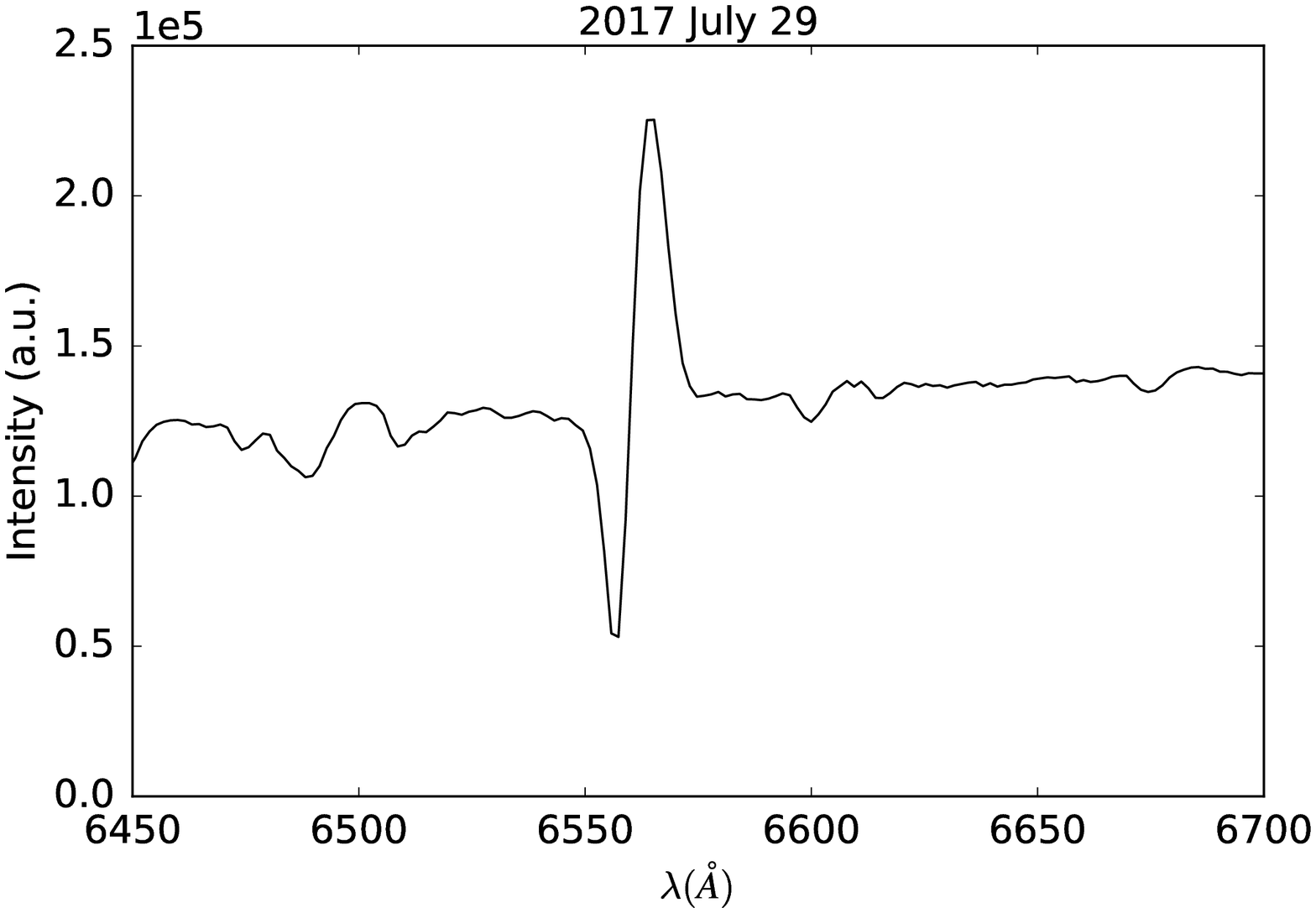}
\caption{Profiles of H$\alpha$ secured during the pre-maximum plateau (left) and close to the maximum (right)}
\label{fig:ha}
\end{figure}

The increasing depth of the H$\alpha$ absorption component from July 12, during the pre-maximum plateau, to July 29, around the maximum, is reported in Fig. \ref{fig:ha}. V723 Cas showed a similar evolution of absorption components during the rise to the main maximum \cite{Iijima1998}.

The observations of ASASSN-17hx show the features of both He/N and Fe II novae at different epochs. Since the classification by \cite{Williams1991}, \cite{Williams1992}, \cite{Williams1994} deals with post-outburst spectra only, ASASSN-17hx joins novae V5558 Sgr \cite{Tanaka2011b} and T Pyx \cite{Imamura2012}, \cite{Ederoclite2014}, confirming that  most novae show both types of spectra during their evolution \cite{Williams2012}.

\begin{figure}
\centering
\includegraphics[width=.4\textwidth]{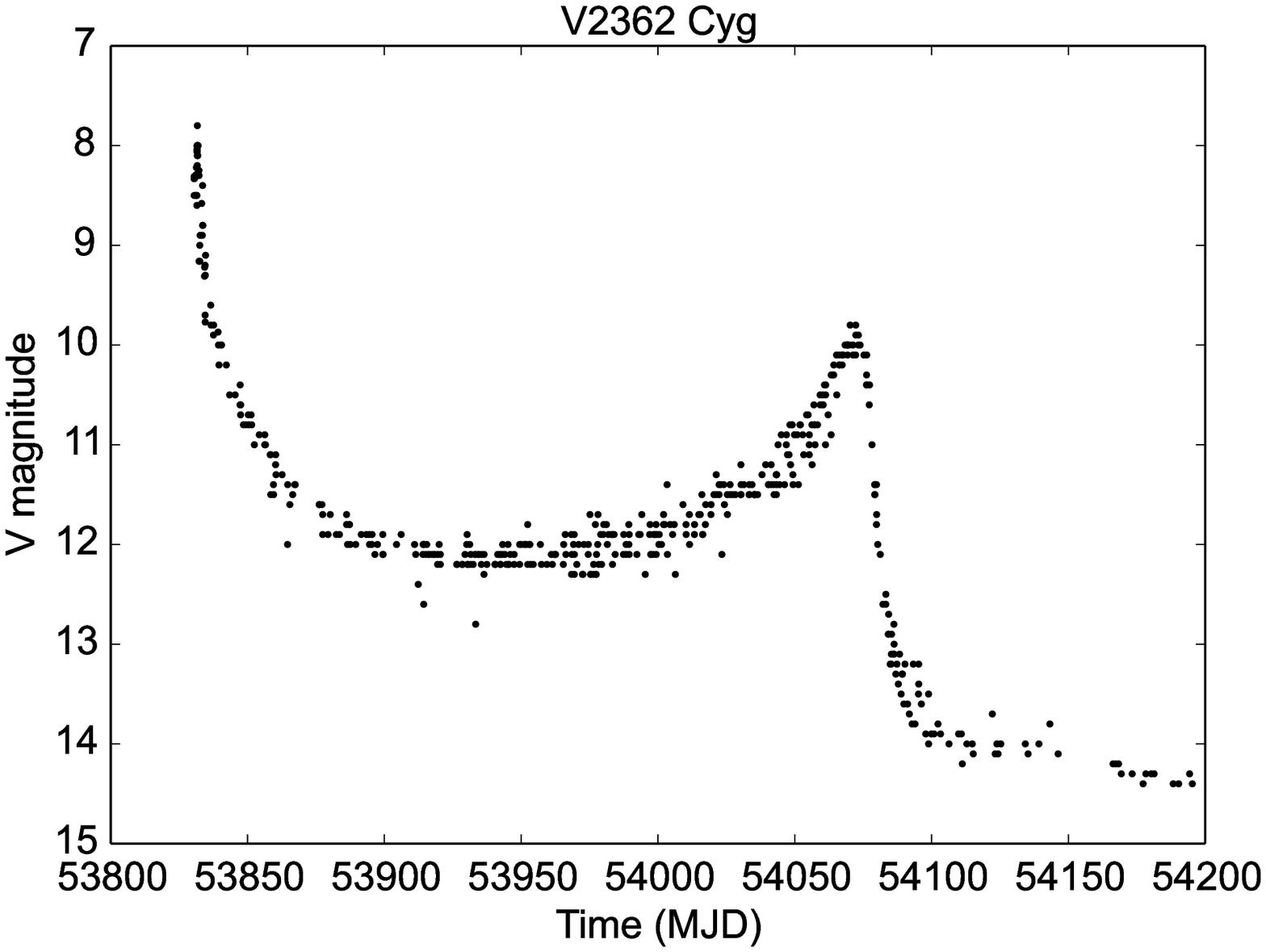}
\includegraphics[width=.4\textwidth]{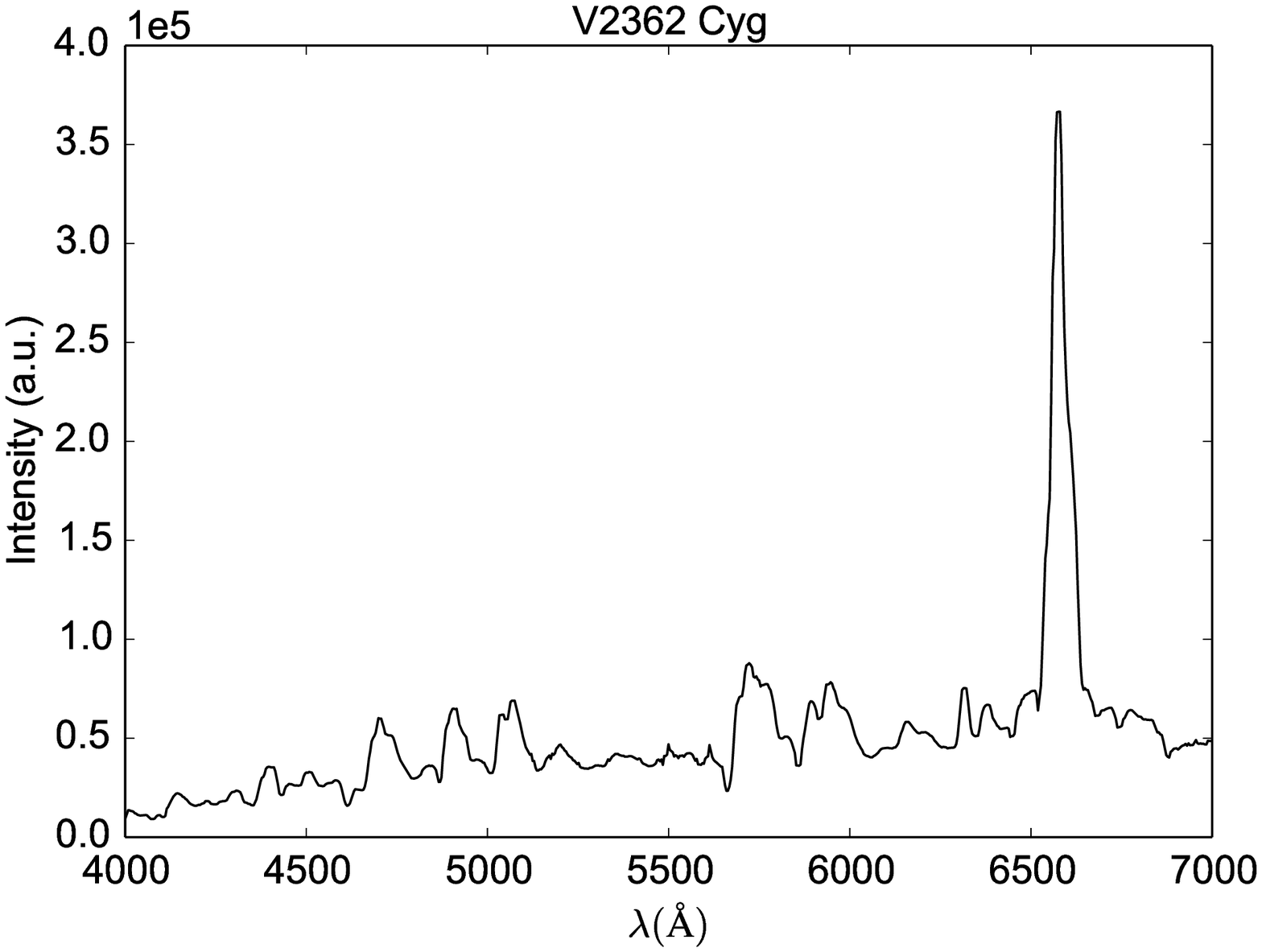}
\caption{Light curve of V2362 Cyg (left) and spectrum secured during the rising phase of the rebrightening (right)}
\label{fig:cusp}
\end{figure}

The evolution of ASASSN-17hx after the main maximum has shown a rebrightening with an amplitude approaching that of the first maximum \cite{Munari2017c}, \cite{Kurtenkov2017b}, with Balmer and Fe II lines with P Cyg absorptions, as in Fe II novae, before the reappearance of He I \cite{Guarro2017}. The spectra reverted to spectra typical of the initial decline during the rise to the second maximum. The evolution of the novae V1186 Sco,   V2540 Oph,   V5113 Sgr, V4745 Sgr, V458 Vul and V378 Ser, that showed rebrightnenings during their decline have been studied by \cite{Tanaka2011a}. The spectra of the above novae showed lines P Cyg profiles at the rebrightenings, but not in the interval between the peaks. The spectra of V5558 Sgr at the rebrightenings showed lines with P Cyg profiles \cite{Tanaka2011b}. The presence of P Cyg lines has been osserved during the rebrightenings of cusp novae \cite{Strope2010} as V1493  Aql,  V2491  Cyg,  V2362  Cyg, V868 Cen, that show a large amplitude secondary peak, explained by a second mass ejection. An example is given by V2362 Cyg \cite{Poggiani2009} (Fig. \ref{fig:cusp}).

\section*{Conclusions}

ASASSN-17hx  is  a  slow  nova  as  V723  Cas  and  V5558  Sgr, showing rebrigthenings during the decline stage. The probable white dwarf mass, 0.6 M$_{\odot}$, is at the limit of thermonuclear runaway. The  pre-maximum  and maximum  spectra  of ASASSN-17hx  show  Fe  II  lines with P Cyg profiles, while the spectrum around the maximum is typical of a supergiant. The  spectral  evolution  during  the  pre-maximum  and  the  early  decline  is similar to those of V5558 Sgr and V723 Cas at the same stage.

\acknowledgments

The author is grateful to the organizers for the invitation to the workshop. Many thanks to Ivan Bruni and to the Telescope Allocation Time of the Loiano Observatory for the observing time. Thanks to ASASSN and to the VSNET, AAVSO observers.


\begin{thebibliography}{99}
\bibitem{asas} http://www.astronomy.ohio-state.edu/asassn/index.shtml
\bibitem{Berardi2017} P. Berardi et al., \emph{ATel} \textbf{10558} (2017).
\bibitem{BodeEvans2008} M. F. Bode and A. Evans, \emph{Classical Novae}, Cambridge University Press (2012).
\bibitem{Cao2012} Y. Cao et al., \emph{ApJ} \textbf{752} (2012) 133. 
\bibitem{Ederoclite2014} A. Ederoclite, \emph{ASPC} \textbf{490} (2014) 57.
\bibitem{Eyres2017} S. P. S. Eyres et al., \emph{MNRAS} \textbf{467} (2017) 2684.
\bibitem{Friedjung1992} M Friedjung, \emph{A\&A} \textbf{262} (1992) 487.
\bibitem{Guarro2017} J. Guarro et al., \emph{ATel} \textbf{10737} (2017).
\bibitem{Hachisu2004} I. Hachisu and M. Kato, \emph{AJ} \textbf{612} (2004) L57.
\bibitem{Hillman2014} Y. Hillman et al., \emph{MNRAS} \textbf{437} (2014) 1962.
\bibitem{Hounsell2010} R. Hounsell et al., \emph{ApJ} {\bf 734} (2010) 480.
\bibitem{Hounsell2016} R. Hounsell et al., \emph{ApJ} {\bf 820} (2016) 104.
\bibitem{Iijima1998} T. Iijima et al., \emph{A\&A} \textbf{338} (1998) 1006. 
\bibitem{Imamura2012} K. Imamura and K. Tanabe, \emph{PASJ} \textbf{64} (2012) 120.
\bibitem{Kato2001} T. Kato and K. Takamizawa, \emph{IBVS} \textbf{5100} (2001).
\bibitem{Kato2002} T. Kato et al., \emph{PASJ} \textbf{54} (2002) 1009.
\bibitem{Kuin2017} N. P. M. Kuin et al., \emph{ATel} \textbf{10636}  (2017).
\bibitem{Kurtenkov2017} A. Kurtenkov et al., \emph{ATel} \textbf{10527} (2017).
\bibitem{Kurtenkov2017b} A. Kurtenkov et al., \emph{ATel} \textbf{10725} (2017).
\bibitem{McLaughlin} D. B. McLaughlin, \emph{ATel} \textbf{10725} (2017).
\bibitem{Munari2017a} U. Munari et al., \emph{ATel} \textbf{10572} (2017).
\bibitem{Munari2017b} U. Munari et al., \emph{ATel} \textbf{10641} (2017).
\bibitem{Munari2017c} U. Munari et al., \emph{ATel} \textbf{10736} (2017).
\bibitem{Orio1993} M. Orio and G. Shaviv, \emph{Ap\&SS} \textbf{2002} (1993) 273.
\bibitem{Pavana2017} M. Pavana et al., \emph{ATel} \textbf{10613} (2017).
\bibitem{PayneGaposchkin1957} C. H. Payne-Gaposchkin, \emph{The Galactic Novae}, North-Holland Pub. Co. (1957).
\bibitem{Pietsch2007a} W. Pietsch et al., \emph{ATel} \textbf{1149} (2007).
\bibitem{Pietsch2007b} W. Pietsch et al., \emph{ATel} \textbf{1157} (2007).
\bibitem{Poggiani2008} R. Poggiani, \emph{NewA} \textbf{13} (2008) 557.
\bibitem{Poggiani2009} R. Poggiani, \emph{NewA} \textbf{14} (2009) 4.
\bibitem{Poggiani2012} R. Poggiani, \emph{MmSAI} \textbf{83} (2012) 753.
\bibitem{Prialnik1995} D. Prialnik and A. Kovetz, \emph{ApJ} \textbf{445} (1995) 789.
\bibitem{Saito2017} R. K. Saito et al., \emph{ATel} \textbf{10552} (2017).
\bibitem{Stanek2017} K. Z. Stanek et al., \emph{ATel} \textbf{10523} (2017).
\bibitem{Strope2010} R. J. Strope et al., \emph{AJ} {\bf 140} (2010) 34.
\bibitem{Tanaka2011a} J. Tanaka et al., \emph{PASJ} \textbf{63} (2011) 159.
\bibitem{Tanaka2011b} J. Tanaka et al., \emph{PASJ} \textbf{63} (2011) 911.
\bibitem{Thompson2017} W. T. Thompson, \emph{MNRAS} \textbf{470} (2017) 4061.
\bibitem{Williams1991} R. E. Williams et al., \emph{ApJ} {\bf 376} (1991) 721.
\bibitem{Williams1992} R. E. Williams, \emph{AJ} {\bf 104} (1992) 725.
\bibitem{Williams1994} R. E. Williams et al., \emph{ApJSS} {\bf 90} (1994) 297.
\bibitem{Williams2012} R. Williams, \emph{AJ} \textbf{144} (2012) 98.
\bibitem{Williams2017} S. C. Williams and M. J. Darnley, \emph{ATel} \textbf{10542} (2017).
\end{thebibliography}
\end{document}